\documentclass[
    ,final            
  ]
  {aipproc}

\layoutstyle{6x9}

\newcommand{\smyr}      {{ M_\odot\ \rm yr^{-1}}}
\newcommand{\sm}        {{ M_\odot}}

\newcommand{\beq}       {\begin{equation}}
\newcommand{\eeq}       {\end{equation}}
\newcommand{\beqa}      {\begin{eqnarray}}
\newcommand{\eeqa}      {\end{eqnarray}}

\newcommand{\mdsd}      {\dot m_{*d}}

\newcommand{\esdb}       {\bar\epsilon_{*d}}
\newcommand{\fsh}       {f_{\rm sh}}

\newcommand{\msdt}      {m_{*d,\,2}}

\newcommand{\teff}      {T_{\rm eff}}

\def\lesssim{\mathrel{\hbox{\rlap{\hbox{\lower4pt\hbox{$\sim$}}}\hbox{$<$}}}}
\def\gtrsim{\mathrel{\hbox{\rlap{\hbox{\lower4pt\hbox{$\sim$}}}\hbox{$>$}}}}

\newcommand{\apj}       {\emph{ApJ}}

\begin{document}

\title{Protostellar Feedback Processes\\ and the Mass of the First Stars}



\author{Jonathan C. Tan}{
  address={Depts. of Astronomy \& Physics, Univ. of Florida, Gainesville, FL 32611, USA. jt@astro.ufl.edu}
}

\author{Britton D. Smith}{
  address={Dept. of Astrophysical \& Planetary Sciences, Univ. of Colorado, Boulder, CO 80309, USA.}
}

\author{Brian W. O'Shea}{
  address={Dept. of Physics \& Astronomy, Michigan State Univ., East Lansing, MI 48864, USA.}
}


\begin{abstract}
We review theoretical models of Population III.1 star formation, focusing on the protostellar feedback processes that are expected to terminate accretion and thus set the mass of these stars. We discuss how dark matter annihilation may modify this standard feedback scenario. Then, under the assumption that dark matter annihilation is unimportant, we predict the mass of stars forming in 12 cosmological minihalos produced in independent numerical simulations. This allows us to make a simple estimate of the Pop III.1 initial mass function and how it may evolve with redshift.
\end{abstract}

\maketitle


\section{Introduction}

The first stars laid the foundations for galaxy formation. As described
below, we expect that these Population III.1 stars, i.e. having
negligible metallicity and experiencing negligible influence from other
astrophysical sources external to their minihalos, were massive
and thus injected significant radiative, mechanical and chemical
feedback into their surroundings. We need to understand the processes
that set the mass of these stars in order to have a complete theory of
galaxy formation and evolution. Supermassive black holes reside in the
centers of most large galaxies and some massive examples have been seen at
very early times \cite{fan2003,willott2003}. The remnants of the first
stars may have been the seeds of these black holes.

We first review the accretion and feedback processes that are expected to occur
during the birth of Pop III.1 stars. We emphasize some of the
uncertainties in existing models. We then describe preliminary
work applying these models to a population of Pop III.1 minihalos in
order to predict the initial mass function (IMF) of the stars they produce.

\section{Accretion and Feedback Processes}

Here we give a brief outline of these processes, which have been
reviewed in more detail by Tan \& McKee\cite{tan2008}. Small dark
matter halos grow and merge together to eventually form $\sim
10^6M_\odot$ ``mini-halos'' by redshifts $z \sim 30-20$. Baryons
collect in these halos and cool down to $\sim 200$~K due
to the presence of trace amounts of $\rm H_2$, the formation of which
is catalyzed by small concentrations of free electrons
via the $\rm H^-$ channel. The baryons start to dominate the mass
density in the innermost $\sim1$ parsec. Above the critical density of
$n_{\rm H}\sim 10^4\:{\rm cm^{-3}}$, cooling is inefficient,
contraction is slow and subsonic, and the temperature gradually
rises towards the center. The gas develops a self-similar, power law
density structure with $\rho \propto r^{-2.2}$, i.e. slightly steeper
than a singular, isothermal sphere. When the central density exceeds
$\sim 10^{10}\:{\rm cm^{-3}}$, the gas becomes fully molecular via the
3-body formation channel, cooling is much more efficient and dynamical
collapse is initiated from the inside-out.

An analytic estimate of the accretion rate to the central star$+$disk system is
\cite{tan2004}:
\beq
\label{eq:mdot}
\mdsd=0.026 \epsilon_{*d}K'^{15/7} (M/\sm)^{-3/7}~M_\odot~{\rm yr}^{-1},
\eeq
where $M$ is the mass originally enclosed by the currently accreting mass shell, $\epsilon_{*d}$ is the instantaneous star formation efficiency, and $K'$ is the ``entropy parameter'' normalizing the polytropic description of the initial core, $P=K\rho^\gamma$ with $\gamma\simeq 1.1$, and is defined as
\beq
K'\equiv \frac{P/\rho^\gamma}{1.88\times 10^{12}~{\rm cgs}}=
         \frac{\teff'}{300~{\rm K}}\left(\frac{10^4{\rm cm}^{-3}}{n_{\rm H}}\right)^{0.1},
\label{eq:kp}
\eeq
where 
$\teff'$
is an effective
temperature that includes the modest effect of subsonic turbulent motions
that are seen in the numerical simulations of Abel, Bryan \& Norman\cite{abn2002}.

The gas core has significant rotational motions\cite{abn2002} and so a
disk should form inside the sonic point of the accretion flow.  We
expect the disk mass to build up until gravitational instabilities
produce spiral density waves and other structures that
allow transfer of mass inwards and angular momentum outwards through
the disk. For fiducial accretion rates and disk sizes, we expect the
optically thick region of the disk to be stable to gravitational
fragmentation\cite{tan2004b}. Most simulations of Pop III.1 star
formation appear to produce single stars\cite{turk2009}, although
these have followed only the earliest stages of accretion when
the central protostar has yet to accrete most of its expected final
mass.

The accretion disk is the first place where turbulent eddies have time
to amplify significantly seed magnetic field produced by the
Biermann battery mechanism during the formation of the
halo\cite{tan2004b}. If such an accretion disk dynamo produces
large-scale, dynamically-strong B-fields, then a bipolar
magneto-centifugally-driven outflow will be launched, which can reduce
$\epsilon_{*d}$. Tan \& Blackman\cite{tan2004b} estimated
$\epsilon_{*d}\simeq 0.5$ by the time $m_*\sim 200 M_\odot$, which is
not as important as radiative feedback in the fiducial case with
$K'=1$ (see below).

Matter joins the protostar via an accretion boundary layer shock and
the internal energy of the gas that is advected into the protostar is
set by the properties of this shock, the accretion rate and the amount
of energy that is radiated. The internal energy of the accreted gas
in turn helps set the size of the protostar\cite{stahler1986,tan2004}.

Weakly interacting massive particle (WIMP) dark matter may
self-annihilate and provide significant energy input into the
protostar\cite{spolyar2008,natarajan2009}, stabilizing its initial
radius at $\sim 10$~AU, rather than $\sim0.1$~AU in the fiducial
case. The subsequent evolution is uncertain\cite{spolyar2009}, but
contraction to the main sequence may be delayed with respect to the
fiducial case without such support (see below), affecting the
mass scale at which radiative feedback truncates
accretion. Uncertainties include the assumed central density profile
of dark matter and how this may be affected by supersonic baryonic
infall and disk accretion, including heating by spiral density
waves. For the remainder of this article, we consider the case of
negligible WIMP annihilation heating.

The protostellar evolution can be followed given knowledge of the
accretion rate\cite{tan2004,hosokawa2009}. Once the protostar is older
than its local Kelvin-Helmholz time, it starts contracting
towards the main sequence, typically settling there by $\sim
100\:M_\odot$. The star could continue to accrete in this state if
feedback is not strong enough to reverse accretion. Given the
accretion rates of equation (\ref{eq:mdot}) and stellar evolution
timescales of Schaerer\cite{schaerer2002}, the maximum accreted
stellar mass before supernova explosion is $m_{\rm *,max}\simeq 1900 K^{\prime1.28}M_\odot$\cite{mckee2008}.

McKee \& Tan\cite{mckee2008} considered a number of feedback processes
that will intervene before a supernova occurs. Destruction of $\rm
H_2$ coolant in the minihalo by FUV photons was not found to reduce
the accretion rate to the star significantly, since by the time this
occurs the protostar has already developed a potential well deep enough for
atomic cooling to allow accretion. Lyman-$\alpha$ radiation pressure
was also found to be ineffective since it requires multiple
scatterings to build up dynamically important pressures, but this
leads to diffusion and escape of the photons along bipolar
cavities. 

Direct ionization of the accretion flow to create a $\sim 25,000$~K
HII region whose high thermal pressure can reverse accretion is the
first important feedback process, occurring by $\sim$50~$M_\odot$ in
the fiducial case. The HII region quickly breaks out in all directions
above and below the disk, reducing $\epsilon_{*d}$ by factors of
several. By the time $m_*=100\:M_\odot$ the accretion rate from
disk-shielded directions is about a factor of 10 smaller
than in the no-feedback case. These estimates use a 1D model for the
disk vertical thickness and assume gas from disk-shielded directions
accretes as it would in the absence of feedback.

The protostar ionizes the accretion disk atmosphere, which redirects
some ionizing flux to outer parts of the disk that are neutral. An
ionized, ``photoevaporative'' flow develops from these outer regions
with mass-loss rate\cite{hollenbach1994}
\beq
\dot{m}_{\rm evap} 
\simeq 4.1\times 10^{-5} S_{\rm 49}^{1/2}T_4^{0.4}\msdt^{1/2}~~\smyr,
\eeq
where $S_{\rm 49}$ is the H-ionizing photon luminosity in units of
$10^{49}$ photons~$\rm s^{-1}$, $T_4$ is the ionized gas temperature
in units of $10^4$~K, and the star+disk mass has been normalized to
$100\:M_\odot$. This analytic estimate has not
yet been confirmed by numerical simulation.

The final accreted mass, $m_{*f}$, is estimated by the
condition $\dot{m}_{\rm evap}=\dot{m}_{*d}$, yielding\cite{mckee2008}:
\beq
\label{eq:maxm}
m_{*f} = 145\, K'^{60/47}  (2.5/T_4)^{0.24} (\fsh/0.2)^{28/47} (\esdb/0.25)^{12/47} M_\odot,
\eeq
where $\fsh$ is the shadowing factor, i.e. fraction of the sky seen
from the protostar blocked by the disk, and $\esdb$ is the
mass-averaged value of $\epsilon_{*d}$ over the formation of the star.
The fiducial normalizations of these parameters were derived from
calculations of the protostellar and accretion disk evolution coupled
with the effects of ionizing feedback. The predicted mass of Pop III.1
stars in the fiducial case is $\sim150\:M_\odot$. The strongest
parameter dependence is with $K^\prime$, i.e. via the accretion
rate. High accretion rate protostars reach higher masses because
photoevaporative truncation occurs at a higher mass, partly because
contraction to the main sequence and HII region breakout were delayed,
thus raising $\esdb$, and partly because the star must produce more
ionizing photons to increase photoevaporative mass-loss
to counter the higher accretion rate.

A number of factors are likely to raise this estimate of the final
accreted mass, including instabilities during the breakout of the HII
region and rotation of the protostar that reduces the equatorial flux
of ionizing photons and thus the photoevaporative mass-loss.

\section{The Pop III.1 IMF}

O'Shea \& Norman \cite{oshea2007} studied the properties of Pop III.1
pre-stellar cores as a function of redshift. They found that cores at
higher redshift are hotter in their outer regions, have higher free
electron fractions and so form larger amounts of $\rm H_2$ (via $\rm
H^-$), although these are always small fractions of the total mass. As
the centers of the cores contract above the critical density of
$10^4\:{\rm cm^{-3}}$, those with higher $\rm H_2$ fractions are able
to cool more effectively and thus maintain lower temperatures to the
point of protostar formation. The protostar thus accretes from
lower-temperature gas and the accretion rates, proportional to $c_s^3
\propto T^{3/2}$, are smaller. Measuring infall rates at the time of
protostar formation at the scale of $M=100\sm$, O'Shea \& Norman found
accretion rates of $\sim 10^{-4}\smyr$ at $z=30$, rising to $\sim
2\times 10^{-2}\smyr$ at $z=20$. They used these accretion rates to
estimate the mass of the star that forms by finding the condition of
when the accretion timescale was equal to the Kelvin-Helmholtz
timescale, thus predicting an increase in the typical Pop III.1
stellar mass as $z$ decreases over this range.

\begin{figure}
\label{fig:mdot}
\includegraphics[height=5.5in]{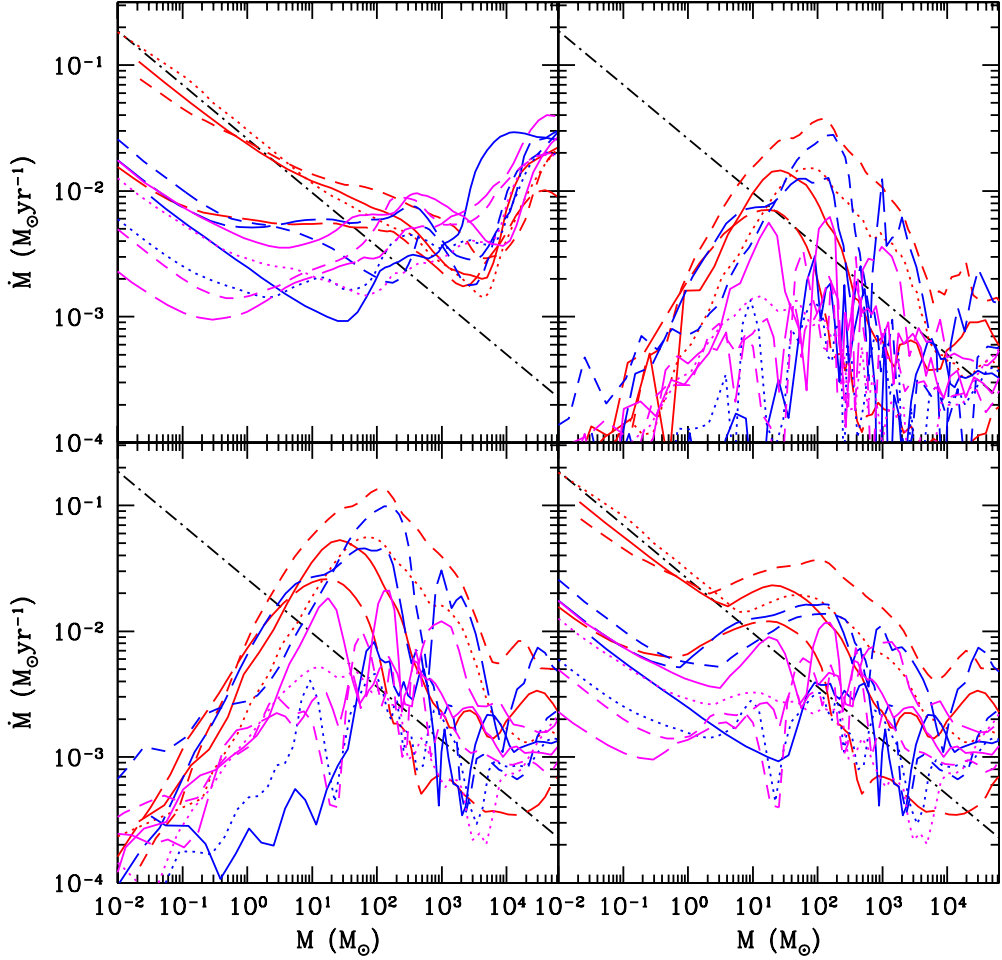}
\caption{Estimates of mass accretion rates of the 12 Pop III.1 cores of O'Shea \& Norman\cite{oshea2007}, each indicated by a different line style and color, with fiducial rate of Tan \& McKee\cite{tan2004} (eq. 1) shown by the black dot-dashed line: (a) Top left: Accretion rate based on SPS collapse with $K^\prime$ profiles (averaged over interior mass) (eq.~\ref{eq:mdot}); (b) Top right: Accretion rate based on mass flux through spherical shells at the end of the simulations; (c) Bottom left: Accretion rate as in (b), now smoothing at 50\% weight with the two adjacent mass shells and an enhancement by a factor of 3.7 as expected if the Hunter\cite{hunter1977} solution holds; (d) Bottom right: Final adopted accretion rate, assuming inner SPS solution (a), then geometric mean of the (a) and (c) results when the latter starts to dominate, then (c) if (a) starts to exceed (c) again (see text).}
\end{figure}

Here we present preliminary results of a more detailed estimate of the
Pop III.1 stellar masses that result from these cores. First we
estimate the expected protostellar accretion rates in the O'Shea \&
Norman\cite{oshea2007} cores (Fig.~\ref{fig:mdot}). In the
simulations, the inner regions have not yet reached the densities
associated with dynamical collapse, so the observed mass infall rate
underestimates the expected accretion rates. Thus, we assume the
solution describing the collapse of a singular polytropic sphere
(SPS)\cite{tan2004} holds for the inner mass shells. We derive this by
evaluating the interior mass averaged $K^\prime$ profiles of the
simulated cores and then using eq.~\ref{eq:mdot}. The infall mass flux
through spherical shells provides another method of estimating the
eventual accretion rate to the protostar. For Hunter's mildly subsonic
solution \cite{hunter1977} used in eq.~\ref{eq:mdot} the accretion
rate increases from $0.70 c_s^3/G$ at large radii ($r\gg c_s t$, where
$t=0$ is the moment of protostar formation) to $2.58c_s^3/G$ at small
radii, an increase of a factor 3.7. Thus we increase the observed
infall mass flux by this factor, and refer to this as the ``enhanced
infall rate''. Beyond the mass shell where this mass accretion rate
becomes larger than that predicted by the collapse of the SPS, we
assume the accretion rate is equal to their geometric mean. This is
designed to approximate the transition from the SPS solution to the
infall solution. In the outer parts of the core, the gas has not yet
virialized, $K^\prime$ becomes large and the SPS-based accretion rate
becomes unrealistically large, exceeding the enhanced infall rate. In
this case, we adopt only the enhanced infall rate as the estimate of the
mass accretion rate.

We then estimate the final protostellar mass, $m_{*,f}$, using the
feedback models of McKee \& Tan\cite{mckee2008}. The collapsed mass at
HII region breakout can be evaluated from their
eq.~(41) with parameters: $\phi_{\rm Edd}=0.3$, $\mu=1/\sqrt{2}$,
$f_d=1/3$, $\phi_S=1$, $T_4=2.5$, $f_{\rm Kep}=0.5$, $r_{\rm
HII}=2r_g$ and $\epsilon_{*d}=1$, then $m_{\rm
*,break}=48.4K^{\prime25/14}M_\odot=6.12\dot{m}_{*d,-3}^{35/27}M_\odot$. Following
the accretion history of each halo, we assume that once this condition
is met HII region breakout is fast and $\dot{m}_{*d}$ is
immediately reduced by the disk shadowing factor $f_{\rm
sh}=0.2$. Finally, we calculate when disk
photoevaporation overwhelms accretion, i.e. $\dot{m}_{\rm
evap}\rightarrow 1.96\times 10^{-4} m_{*,2}^{5/4}\:{\rm
M_\odot\:yr^{-1}}>\dot{m}_{*d}$\cite{mckee2008}, at which point we
assume the protostar has reached its $m_{*f}$. A more
detailed analysis, fully coupling $\dot{m}_{*d}$ to these
protostellar evolution and feedback models will be presented in a
future paper.

Figure 2 shows the derived final protostellar masses as a function of
the collapse redshift of the minihalo. There is a weak trend for
lower-mass stars to be formed at higher redshift due to lower
accretion rates, as suggested by O'Shea \&
Norman\cite{oshea2007}. However, there is significant dispersion
about this trend, since although the high $z$ protostars do initially
have low accretion rates, these can increase due to large mass infall
rates at the mass shells for $M\sim 100-1000\:M_\odot$. Figure 2 also
shows the distribution of the initial Pop III.1 stellar masses,
i.e. their IMF, based on these 12 minihalos. These are the first halos
to collapse in their local cosmological volumes, so may be
a biased sample. The sample is also small, so we note only some
basic features: we see very few stars below $80\:M_\odot$, a steep
rise in the IMF around 100~$M_\odot$, then a general decline to higher
masses, with the most massive star being $\sim 600\:M_\odot$. The mean
mass is $\sim 250\:M_\odot$, the median is $\sim 215\:M_\odot$ and the
dispersion is $\sim 160\:M_\odot$.


\begin{figure}
\label{fig:imf}
  \includegraphics[height=2.7in]{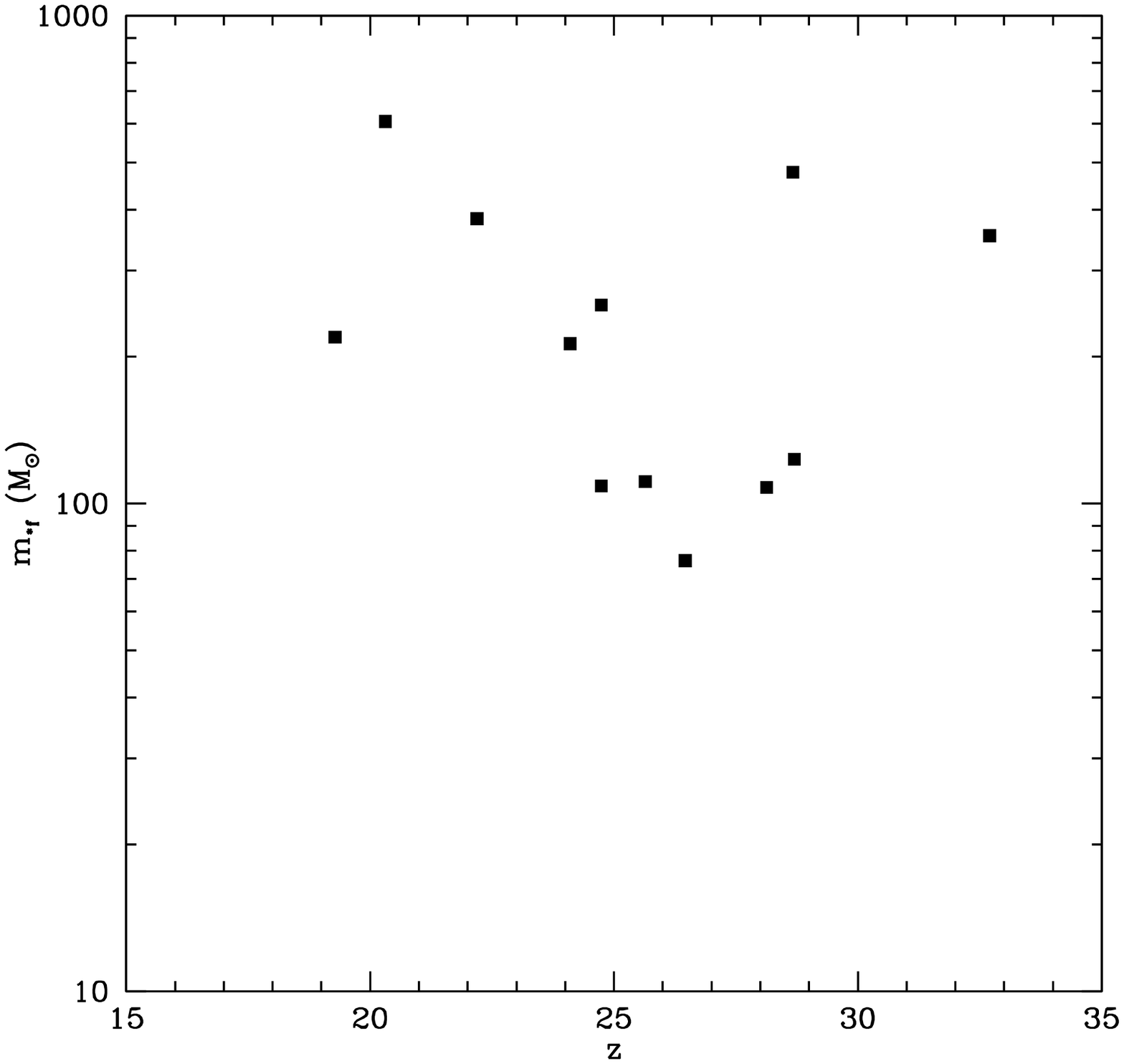}
\includegraphics[height=2.7in]{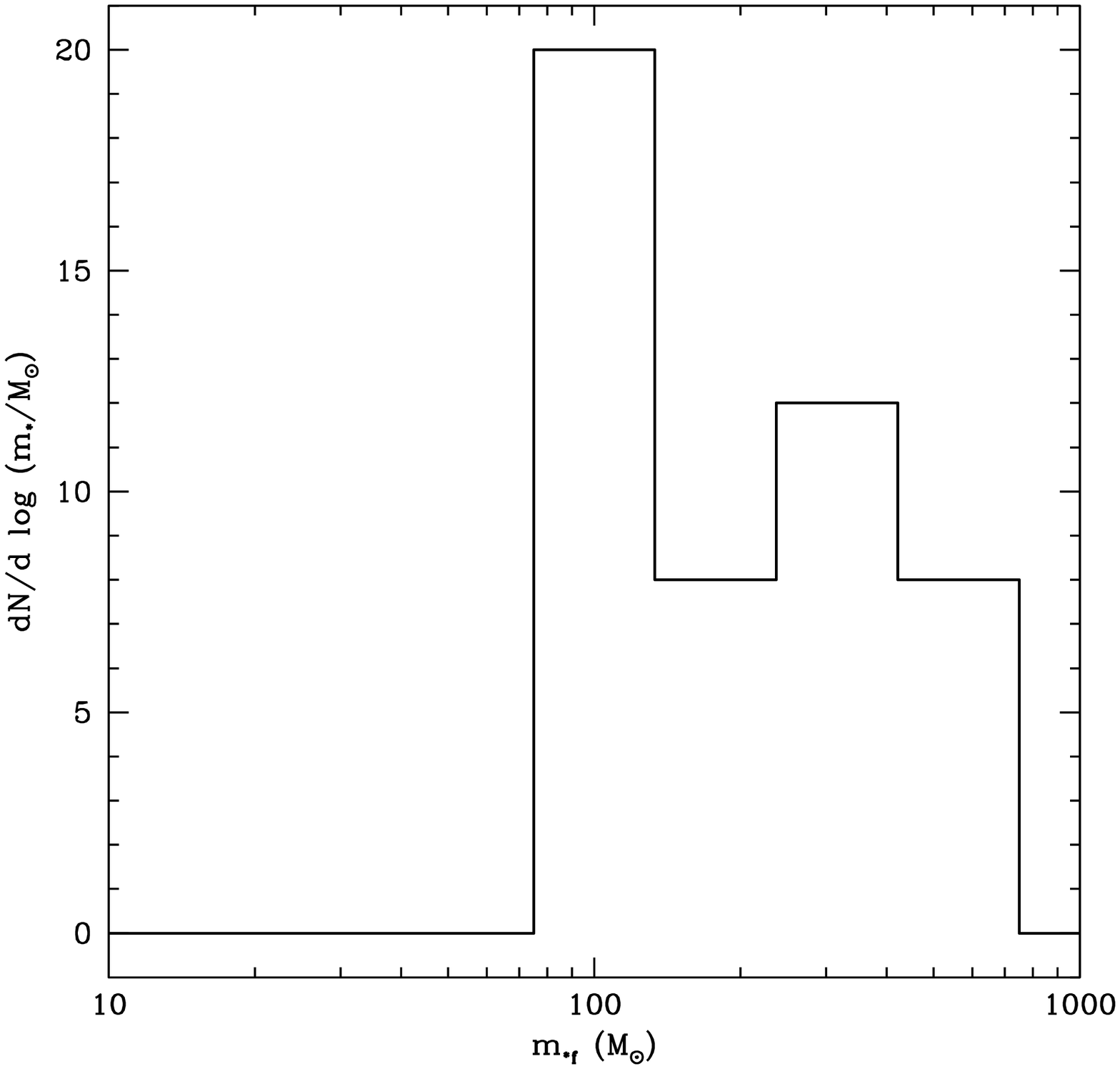}
  \caption{Left: Final protostellar mass due to photoevaporative feedback, $m_{*f}$, versus $z$ for the 12 halos of O'Shea \& Norman\cite{oshea2007}. Right: IMF of the stars forming in these Pop III.1 minihalos.}
\end{figure}



\begin{theacknowledgments}
We acknowledge NSF CAREER grant AST-0645412 (to JCT), the AstroWIN workshop at the Dept. of Astronomy, Univ. of Florida, and thank C. McKee for helpful comments.
\end{theacknowledgments}

\end{document}